\documentclass[prd, preprint, 11pt]{revtex4-1}

\usepackage{amsmath}
\usepackage{amssymb}
\usepackage{setspace}
\usepackage{graphicx}
\usepackage{natbib}
\usepackage{float}
\allowdisplaybreaks

\begin{document}
 
 % ! TEX spellcheck
 %
%\ \vskip 1.0 in

\begin{center}
 { \large {\bf A possible correspondence between Ricci identities and Dirac equations in the Newman-Penrose  formalism  \\ {\it Towards an understanding of gravity induced collapse of the wave-function?}}}

%\smallskip

\vskip 0.2 in

{\large{\bf Anushrut Sharma$^{*}$ and Tejinder P.  Singh$^{\dagger}$}} 

{\it $^{*}$Indian Institute of Technology Bombay, Powai, Mumbai 400076}\\  
{\it $^{\dagger}$Tata Institute of Fundamental Research,}
{\it Homi Bhabha Road, Mumbai 400005, India}\\
{\tt email: anushrut@iitb.ac.in, tpsingh@tifr.res.in}\\
\medskip

%\vskip 0.5cm
\end{center}

\centerline{\bf ABSTRACT}
\noindent It is well-known that in the Newman-Penrose formalism the Riemann tensor can be expressed as a set of eighteen complex first-order equations, in terms of the twelve spin coefficients, known as Ricci identities. The Ricci tensor herein is determined via the Einstein equations. It is also known that the Dirac equation in a curved spacetime can be written in the Newman-Penrose formalism as a set of four first-order coupled equations for the spinor components of the wave-function. In the present article we suggest that it might be possible to think of the Dirac equations in the N-P formalism as a special case of the Ricci identities, after an appropriate identification of the four Dirac spinor components with four of the spin coefficients, {\it provided torsion is included in the connection}, and after a suitable generalization of the energy-momentum tensor. We briefly comment on similarities with the Einstein-Cartan-Sciama-Kibble theory. The motivation for this study is to take some very preliminary steps towards developing a rigorous description of the hypothesis that dynamical collapse of the wave-function during a quantum measurement is caused by gravity.

\smallskip

%\setstretch{1.1}

\noindent 
\medskip

\smallskip

%\setstretch{1.1}

\section{Introduction and motivation}

\noindent One of the possible ways to understand the outcome of a quantum measurement is to propose that there is a dynamical collapse of the wave-function. Some non-linear physical process causes a breakdown of quantum linear superposition, leading to one or the other branches of the superposition being actually realized randomly in a given sampling of the measurement, in accordance with the Born probability rule. A successful phenomenological description of dynamical collapse is the model of Continuous Spontaneous Localization [CSL] wherein the non-relativistic Schr\"{o}dinger equation is modified to include  a non-linear stochastic component \cite{Ghirardi:86}, \cite{Ghirardi2:90}, \cite{Pearle:76}, \cite{Bassi:03}.
CSL successfully explains what is observed during a quantum measurement, and its predictions of departures from quantum theory  are not contradicted by current laboratory tests and astronomical constraints (for a recent review see \cite{RMP:2012}). What is missing though is an infallible understanding  of the physical process responsible for the proposed stochastic modification of the Schr\"{o}dinger equation. 

It has been proposed by at least  three different researchers, Karolyhazy \cite{Karolyhazi:66},  \cite{Karolyhazi:86}, Penrose \cite{Penrose:96} and Diosi \cite{Diosi:87} and their collaborators, that such a physical process could be provided by self-gravity. Although their approaches differ  in detail, the overall physical nature of the process could roughly be described as follows.  A quantum object, while in motion, distorts the geometry of the spacetime around itself, causing it to fluctuate, by virtue of its own gravity. This fluctuating geometry causes loss of phase coherence in the wave-function, and consequent spatial localization of the wave-function, with the effect inevitably becoming more important for objects with higher mass. This has been demonstrated through model calculations by Karolyhazy and collaborators, and similar results have been obtained by Diosi and by Penrose, and by other researchers who have subsequently analyzed their proposal \cite{Frenkel:2002}, \cite{Giulini2011}, \cite{Giulini2012}, \cite{Giulini2013}, (also see \cite{Hu2014}, \cite{Colin2014}). For  recent reviews of gravity induced collapse see \cite{Gao} and \cite{Diosi2013}. For a fascinating different recent proposal as to involvement of gravity in wave-function collapse through a complex metric see \cite{Adler2014}.

Here we follow up on a lead provided by some of Karolyhazy's results, and we suggest some very preliminary  ideas as to how one might proceed, from first principles, towards arriving at a fundamental stochastic nonlinear Schr\"odinger equation which incorporates the effect of self-gravity to explain collapse of the wave-function during measurement and spatial localization of macroscopic objects. Karolyhazy shows that for an elementary particle of mass $m$, the coherence length $a_c$ over which the wave-function of the particle is localized is given by
\begin{equation}
a_c \approx \frac{\hbar^2}{G}\; \frac{1}{m^3}  \approx \left(\frac{L}{l_p}\right)^{2} L; \qquad L= \frac{\hbar}{mc}
\label{micro}
\end{equation} 
For an extended subject of size $R$ the localization length is given by
\begin{equation}
a_c \approx \left(\frac{\hbar^2}{G}\right)^{1/3}\; \frac{R^{2/3}}{m} \approx 
\left(\frac{R}{l_p}\right)^{2/3} L
\label{macro}
\end{equation}
It is shown in \cite{Frenkel:2002} that for a micro-object of linear size $R\ll a_c$ Eqn. (\ref{micro}) continues to hold. [For $a_c=R$ the two expressions coincide]. Thus Eqn. (\ref{macro}) can be taken as the general expression for localization length, and the following important inferences hold:
\begin{equation}  
a_c\gg R \implies \frac{\hbar^2}{G} \gg m^3 R \qquad: micro-region
\end{equation}
\begin{equation}
a_c \approx R \implies \frac{\hbar^2}{G} \approx m^3 R \qquad: transition-region
\end{equation}
\begin{equation}
a_c\ll R \implies \frac{\hbar^2}{G} \ll m^3 R \qquad: macro-region
\end{equation}

Eqn. (\ref{macro}) can be written in the following useful form
\begin{equation}
\frac{a_c}{R} \approx \left(\frac{L}{R_S}\right)^{2/3}\; \left(\frac{R_S}{R}\right)^{1/3}
\label{ratio}
\end{equation}
where $R_S=Gm/c^2$ is of the order of the Schwarzschild radius of the object. We now specialize to the theoretically simpler case where $R=R_S$, i.e. objects whose physical size is of the order of their Schwarzschild radius, and hence we would be dealing only with objects which are described as black holes, {\it if} they are classical objects. Hence Eqn. (\ref{ratio}) becomes
\begin{equation}
\frac{a_c}{R} = \left(\frac{L}{R_S}\right)^{2/3}
\end{equation}
It is clear that in the classical macro-limit $a_c\ll R$ the Schwarzschild radius far exceeds the Compton wavelength as one would expect, and this also implies that $m\gg m_{pl}$, $R_S\gg l_p$ and $L\ll l_p$. The macroscopic black hole limit is obtained for large masses such that Schwarzschild radius far exceeds Planck length, and Compton wavelength is much less than Planck length. Conversely the quantum micro-limit 
$a_c\gg R$ is realized if  Compton wavelength far exceeds Schwarzschild radius, i.e.   $m\ll m_{pl}$, $R_S\ll l_p$ and $L\gg l_p$. The microscopic quantum limit is obtained for small masses such that Schwarzschild radius is far less than Planck length, and Compton wavelength is much larger than Planck length: the concept of black hole is no longer defined in this limit, and one is dealing with a quantum object characterized by its
Compton wavelength.

The macro-limit is well described by general relativity and the Einstein equations for spacetime curvature described by the Riemann tensor, accompanied by classical equations of motion for the particle. The 
micro-limit is described by the Dirac equation  for a relativistic particle of mass $m$.  In both cases [general relativity and the Dirac equation] the mass $m$ acts as a source term. One could well ask as to how the particle knows whether it should obey Einstein equations or the Dirac equation?! In fact there is no 
information either in Einstein equations or the Dirac equation which would provide an answer to this question. It is only through experience that we know that micro-objects [such as an electron] should be described by the Dirac equation and macro-objects [such as a solar mass rotating black hole] should be described by Kerr solution of
Einstein equations. And then there is the argument presented above, for gravity induced quantum-classical transition, where the behaviour of an object is quantum or classical, depending on whether or not the associated Compton wavelength exceeds the Schwarzschild radius.  It would be significant if there were to be a system of equations describing the dynamics of a mass $m$, irrespective of whether or not this mass satisfies $m\ll m_{pl}$ or $m\gg m_{pl}$, and to which system the Einstein equations and the Dirac equation would be approximations. Such a system of equations would significantly help in understanding the transition region $m\sim m_{pl}$ as well as provide an improved quantitative understanding of gravity induced quantum collapse and localization.  A small step in this direction will be proposed in the following sections.

The possibility of such a structure arising is also suggested by schematically looking at a possible action for the gravitational field and the Dirac field, if both were to be sourced by the same mass $m$. Ignoring technical and related conceptual issues, since this is largely a heuristic argument, one could write the action as
\begin{equation}
S = \frac{c^3}{G} \int d^4x \;\sqrt{-g} R + \hbar\int d^{4}x \; \sqrt{-g}\;\overline{\psi}(x) (i\gamma^{\mu}\partial_\mu\psi) 
- mc \int d^{4}x\; \sqrt{-g}\; \overline{\psi}{\psi}
\label{schematicaction}
\end{equation}
[If one were to imagine getting the Einstein equations out of this action, in the limit $\hbar\rightarrow 0$, then in the last term one could replace $\overline{\psi} \psi$ by a spatial three-delta function 
$\delta^{3}({\bf x})$ representing localization of the mass at a point].

Let us estimate the relative magnitudes of the integrands in these three terms in the action, by introducing characteristic lengths, leaving aside the four volume and metric which are common to all three terms. If there is a characteristic length $l$ associated with the system, the curvature scalar may be estimated via $R\sim 1/l^2$ and the first term is $T_1\sim c^3/Gl^2$, whereas the second term is $T_2\sim \hbar/l^4$ and the third term is $T_3 \sim mc/ l^3$. If $T_1$ dominates over $T_2$, then by virtue of the resulting field equations we expect $T_1 \sim T_3$ and hence
\begin{equation}
\frac{c^3}{G} \frac{1}{l^2} \sim \frac {mc}{l^3} \implies l \sim \frac{Gm}{c^2} \sim  R_S
\end{equation}
If $T_2$ dominates over $T_1$ and is order $T_3$ then
\begin{equation}
\frac{\hbar}{l^4}\sim \frac{mc}{l^3} \implies l \sim \frac{\hbar}{mc} \sim L
\end{equation}
We see that $T_1\gg T_2$  suggests $R_S\gg L$ (the scale implied by $T_2$ should be ignorable) and $T_1\ll T_2$ suggests $R_S \ll L$ (the scale implied by $T_1$ should be ignorable). This indicates that when $T_1\sim T_2$ and $m\sim m_{pl}$ a dynamical description arising out  of a combined consideration of the Dirac action and the Einstein-Hilbert action for a particle of mass $m$ might be possible.

It is interesting that the three terms $T_1, T_2, T_3$ all come with different power-law dependence on the characteristic length scale, being $1/l^2, 1/l^4, 1/l^3$ respectively. As a result, we can write the three integrands together in the following form, after pulling out the constant $c^3/G$:
\begin{equation}
 \frac{G}{c^3} S\sim \int d^4x \sqrt{-g} \; \left[ \frac{1}{l^2}  + \frac{l_p^{2}}{l^4} - \frac{R_S}{l^3} \right]
 \end{equation} 
 This permits thinking of the Dirac field as a modification arising if Planck length is taken as non-vanishing.

In order to make preliminary progress towards describing Einstein equations and Dirac equations as part of a common system, we must first try and find a common language for describing gravity and the Dirac field. Of course gravity described via the metric and the Dirac field described by spinors look very different. On the other hand, more promising is a consideration of Einstein gravity in the tetrad formalism; in particular through the spin coefficents which are used to write down the Riemann tensor in the Newman-Penrose formalism, via the so-called Ricci identities. As for the Dirac equation, this too can be written for a curved space in the Newman-Penrose formalism, and then the set of four Dirac equations look strikingly similar to the Riemann tensor equations [the Ricci identities] written in term of spin coefficients in the N-P language. We then make the unusual proposal that the Dirac spinor components  be identified with four of the spin coefficients.  This allows the four Dirac equations to be deduced from the eighteen complex N-P equations for the Riemann tensor, provided one includes torsion in the theory. This brings in some similarity of the present analysis with the Einstein-Cartan theory, which generalizes general relativity to include torsion. Also, it appears necessary to assume a new form for the energy-momentum tensor, which relates to the Ricci tensor and Weyl tensor via a complex generaization of Einstein equations. Since the Dirac equations are in this manner found to be a special case of the Newman-Penrose equations, this raises the hope that a Dirac equation which incorporates the effect of self-gravity (and whose non-relativistic limit might be of help in understanding gravity induced collapse) might be obtainable from the Ricci identities with torsion, written in the N-P formalism. 

\section{Einstein equations in the Newman-Penrose formalism}
The Newman-Penrose formalism is a tetrad formalism in which the basis vectors are a tetrad of null vectors $\bf l, n, m, \overline{m}$ with ${\bf l}$ and ${\bf n}$ real, and ${\bf m}$ and ${\bf \overline{m}}$ complex conjugates of each other.  These basis vectors can be considered as directional derivatives and are denoted by the special symbols
\begin{equation}
D={\bf l}, \quad \Delta={\bf n}, \quad \delta = {\bf m}, \quad \delta^{*} = \overline{\bf m}
\end{equation}
The Ricci rotation coefficients, also known as spin coefficients, arise in the definition of the covariant derivatives of the four null vectors. There are twelve spin coefficients, denoted by special standard symbols
\begin{equation}
 \kappa, \sigma, \lambda, \nu, \rho, \mu, \tau, \pi, \epsilon, \gamma, \alpha, \beta
 \end{equation} 

The ten independent components of the Weyl tensor are expressed by five complex Weyl scalars, denoted as
\begin{equation}
 \Psi_0, \Psi_1, \Psi_2, \Psi_3, \Psi_4
 \end{equation}
 The ten components of the Ricci tensor are defined in terms of four real scalars and three complex scalars
 \begin{equation}
 \Phi_{00}, \Phi_{22}, \Phi_{02}, \Phi_{20}, \Phi_{11}, \Phi_{01}, \Phi_{10}, \Lambda, \Phi_{12}, \Phi_{21}
 \label{Ricci}
 \end{equation}
  The definitions of these scalars can be found for instance in Chandrasekhar's book {\it The Mathematical Theory of Black Holes} \cite{Chandra}.
  
 The Riemann tensor can be expressed in terms of Weyl scalars and Ricci scalars, and directional derivatives of the spin coefficients. There are eighteen complex equations to this effect, known as Ricci identities, not all independent.  These central equations of the N-P formalism are given by Equations 310 (a-r) of Chapter 1 of Chandrasekhar's book, and we reproduce them below, for our further use.
 \begin{subequations}
 \begin{equation}
Ch. 310 (a) : \quad  D\rho - \delta^{*}\kappa = (\rho^2+\sigma\sigma^{*}) + \rho( \epsilon + \epsilon^{*})
 -\kappa^{*}\tau -\kappa (3\alpha +\beta^{*}-\pi) + \Phi_{00}
 \end{equation}
 \begin{equation}
 Ch. 310 (b) : \quad D\sigma - \delta\kappa = \sigma(\rho + \rho^{*} + 3\epsilon - \epsilon^{*})
 -\kappa(\tau - \pi^{*} + \alpha^{*} +3\beta) + \Psi_{0} 
 \end{equation}
 \begin{equation}
 Ch. 310 (c) : \quad D\tau - \Delta\kappa = \rho(\tau + \pi^{*}) + \sigma (\tau^{*} + \pi) + 
 \tau (\epsilon - \epsilon^{*}) - \kappa (3 \gamma + \gamma^{*}) + \Psi_1 + \Phi_{01}
 \end{equation}
 \begin{equation}
 Ch. 310 (d) : \quad D\alpha -\delta^{*}\epsilon = \alpha (\rho + \epsilon^{*} - 2 \epsilon) + \beta\sigma^{*}
  - \beta^{*}\epsilon - \kappa \lambda - \kappa^{*}\gamma + \pi (\epsilon + \rho) + \Phi_{10}
 \end{equation}
 \begin{equation}
 Ch. 310 (e) : \quad D\beta - \delta\epsilon = \sigma (\alpha + \pi) + \beta (\rho^{*} - \epsilon^{*}) 
 - \kappa ( \mu + \gamma) - \epsilon (\alpha^{*} - \pi^{*}) + \Psi_{1}
 \end{equation}
 \begin{equation}
 Ch. 310 (f) : \quad D\gamma - \Delta\epsilon = \alpha (\tau + \pi^{*}) + \beta (\tau^{*} + \pi) 
 -\gamma (\epsilon + \epsilon^{*}) - \epsilon (\gamma + \gamma^{*}) + \tau\pi - \nu\kappa
 + \Psi_2 + \Phi_{11} - \Lambda 
 \end{equation}
 \begin{equation}
 Ch. 310 (g) \quad D\lambda - \delta^{*}\pi = (\rho\lambda + \sigma^{*}\mu) + \pi (\pi + \alpha - \beta)
 -\nu\kappa^{*} - \lambda ( 3\epsilon - \epsilon^{*}) + \Phi_{20}
 \end{equation}
 \begin{equation}
 Ch. 310 (h) \quad D\mu - \delta\pi = (\rho^{*}\mu + \sigma\lambda) + \pi ( \pi^{*} - \alpha^{*} + \beta)
 - \mu (\epsilon + \epsilon^{*}) - \nu\kappa + \Psi_{2} + 2\Lambda
 \end{equation}
 \begin{equation}
 Ch. 310 (i): \quad D\nu - \Delta\pi = \mu (\pi + \tau^{*}) + \lambda (\pi^{*} + \tau) + 
 \pi (\gamma - \gamma^{*}) - \nu (3\epsilon + \epsilon^{*}) + \Psi_3 + \Phi_{21} 
 \end{equation}
 \begin{equation}
Ch. 310 (j): \quad \Delta\lambda - \delta^{*}\nu = - \lambda (\mu + \mu^{*} + 3\gamma - \gamma^{*})
 + \nu (3\alpha + \beta^{*} + \pi - \tau^{*}) - \Psi_4
 \end{equation}
 \begin{equation}
 Ch. 310 (k): \delta \rho - \delta^{*}\sigma = \rho (\alpha^{*} + \beta) - \sigma (3\alpha - \beta^{*})
 + \tau (\rho - \rho^{*}) + \kappa (\mu - \mu^{*}) - \Psi_1 + \Phi_{01}
 \end{equation}
 \begin{equation}
 Ch. 310 (l): \delta\alpha - \delta^{*}\beta = (\mu\rho - \lambda\sigma) + \alpha\alpha^{*}
 + \beta\beta^{*} - 2 \alpha \beta + \gamma (\rho - \rho^{*}) + \epsilon (\mu - \mu^{*}) -
 \Psi_2 + \Phi_{11} + \Lambda 
 \end{equation}
 \begin{equation}
Ch. 310 (m): \delta\lambda - \delta^{*}\mu = \nu (\rho - \rho^{*}) + \pi (\mu - \mu^{*}) + \mu (\alpha + \beta^{*})
 + \lambda (\alpha^{*} - 3 \beta) - \Psi_3 + \Phi_{21}
 \end{equation}
 \begin{equation}
 Ch. 310 (n): \delta\nu - \Delta\mu = (\mu^2+\lambda\lambda^{*}) + \mu (\gamma + \gamma^{*}) - \nu^{*}\pi
 + \nu (\tau - 3 \beta - \alpha^{*}) + \Phi_{22}  
 \end{equation}
 \begin{equation}
 Ch. 310 (o): \delta\gamma - \Delta\beta = \gamma (\tau - \alpha^{*} - \beta) +\mu\tau - \sigma\nu
 -\epsilon\nu^{*} - \beta (\gamma - \gamma^{*} - \mu) + \alpha\lambda^{*} + \Phi_{12}
 \end{equation}
 \begin{equation}
 Ch. 310 (p) : \delta\tau - \Delta\sigma = (\mu\sigma + \lambda^{*}\rho) + 
 \tau (\tau + \beta - \alpha^{*}) - \sigma (3\gamma - \gamma^{*}) - \kappa\nu^{*} + \Phi_{02}
 \end{equation}
 \begin{equation}
Ch. 310 (q) : \Delta \rho - \delta^{*}\tau = - (\rho\mu^{*} + \sigma\lambda) + 
\tau (\beta^{*} - \alpha - \tau^{*})  + \rho (\gamma + \gamma^{*}) + \nu\kappa - \Psi_2 - 2\Lambda  
 \end{equation}
 \begin{equation}
 Ch. 310 (r) : \Delta\alpha - \delta^{*}\gamma = \nu (\rho + \epsilon) - \lambda (\tau + \beta)
 + \alpha ( \gamma^{*} - \mu^{*}) + \gamma (\beta^{*} - \tau^{*}) - \Psi_3
 \end{equation}
 \end{subequations}

In these equations the Ricci components are to be determined from the Einstein equations. Since there are thritysix real equations, and only twenty independent Riemann components, these equations are subject to sixteen constraints, the so-called eliminant conditions.

\section{A proposal for obtaining Dirac equations as a special case of Ricci identities}

The four Dirac equations in the N-P formalism are given by [Chandrasekhar, p. 543] \cite{Chandra}
\begin{equation}
(D+\epsilon - \rho) F_1 + (\delta^{*} + \pi - \alpha) F_2 = i\mu_{*} G_{1}
\label{D1}
\end{equation}
\begin{equation}
(\Delta + \mu - \gamma) F_2 + (\delta + \beta - \tau) F_1 = i\mu_{*} G_{2}
\label{D2}
\end{equation}
\begin{equation}
(D + \epsilon^{*} - \rho^{*}) G_{2} - (\delta + \pi^{*} - \alpha^{*}) G_{1} =i \mu_{*} F_{2}
\label{D3}
\end{equation}
\begin{equation}
(\Delta + \mu^{*} - \gamma^{*}) G_{1} - (\delta^{*} + \beta^{*} - \tau^{*}) G_{2} = i \mu_{*} F_{1}
\label{D4}
\end{equation}
where $F_1, F_2, G_1$ and $G_2$ are Dirac spinor components.

Here, $\mu_{*} = mc/\sqrt{2}\hbar$. The similarity between the formal structure of the  Dirac equations and the Ricci identities is suggestive. In both cases, each equation involves a pair of derivatives of spin-coefficients / Dirac spinors. We will see how under certain assumptions the Ricci identities in the N-P formalism, after a suitable generalization, might contain within themselves, the Dirac equations as a special case. The first generalization we refer to is that the Riemann tensor should be allowed to be complex, as a consequence of the complexity of the spin coefficients [unlike in the standard case where the Riemann tensor is real, even though the spin coefficients are complex]. This complexification of the Riemann tensor does not prevent the assumption of a real flat Minkowski space-time background, because as we will eventually see, the complex components are assumed to arise from a complex non-zero torsion. The second generalization will consist of a new relation between the Riemann tensor and the energy-momentum tensor, which amounts to suggesting a generalized form for the energy-momentum tensor. The principal assumption is that we propose to make a correspondence between the four Dirac spinor components in the N-P formalism, $F_1, F_2, G_1, G_2$, with four of the spin coefficients.  Since a spinor component has length dimensions $L^{-3/2}$ and a spin coefficient has length dimensions $L^{-1}$, they will be related through a constant of length dimension $L^{1/2}$ which we assume to be the square-root of Planck length $l_p$. Upon inspection and comparison it is found that the appropriate identification between the spin coefficients and the Dirac spinor components is
\begin{equation}
F_1 =\frac{1}{\sqrt{l_p}}\; \lambda, \quad F_2=-\frac{1}{\sqrt{l_p}}\;\sigma, \quad G_1=\frac{1}{\sqrt{l_p}}\;\kappa^{*}, \quad G_2=\frac{1}{\sqrt{l_p}}\;\nu^{*}
\label{match}
\end{equation}
%The spin-coefficient $\gamma$ is non-zero but is not independent, and is determined by the above four spin coefficients, as we shall see. 
The remaining eight spin coefficients are set to zero:
\begin{equation}
\rho=\mu=\tau=\pi=\epsilon=\gamma=\alpha=\beta=0
\label{vanishspin}
\end{equation}
The reason why we chose, $\lambda ,  \sigma , \kappa$ and $\nu$, to be the Dirac spinors is that these are the only four spin coefficients which are absent from the Dirac equations. Further, when we take $\kappa^*$ and $\nu^*$ to be the last two Dirac spinors, we get all the appropriate derivatives from the Ricci identities, as are required by the Dirac equation.

There remains an important unanswered question as to how one could be sure that tetrad transformations will not mix up the spin coefficents and make non-zero those spin-coefficients which we have set to zero? We do not have an answer to this at the present stage, and it remains an assumption. However an indicator that our assumption that these four non-zero spin-coefficients might not mix with the other ones comes from an observation relating to the opposite extreme limit, namely black holes in general relativity: we quote from Chandrasekhar [Chapter 1, Section 9, page 63] \cite{Chandra}. ``It is a remarkable fact that the black-hole solutions of general relativity are all of Petrov type D and therefore, enable their analysis in a null tetrad frame in which the spin coefficients $\kappa, \sigma, \lambda$, and $\nu$ and all the Weyl scalars, except $\Psi_{2}$, vanish.'' Remarkably, it is precisely these spin coefficients which are required to be non-vanishing in the microscopic Dirac limit, while the other eight are set to zero. It is as if there is some complementarity between the black hole limit and the Dirac limit; this gives us some confidence in the assumptions made here.   Furthermore, in the context of the torsion dominated case considered below, it is to be noted that on a Minkowski flat spacetime, the connection is antisymmetric and has four independent components \cite{Connor}.

We now examine the eighteen N-P equations to see how the four Dirac equations can be obtained from them. Thus we re-write the Ricci identities under the conditions (\ref{match}) and (\ref{vanishspin}) :
\begin{subequations}
\begin{equation}
-\delta^*\kappa = \sigma\sigma^* + \Phi_{00}
\label{D3b}
\end{equation}
\begin{equation}
D\sigma - \delta\kappa = \Psi_0
\end{equation}
\begin{equation}
-\Delta\kappa = \Psi_1 + \Phi_{01}
\label{D4b}
\end{equation}
\begin{equation}
0 = -\kappa\lambda + \Phi_{10}
\end{equation}
\begin{equation}
0 = \Psi_1
\end{equation}
\begin{equation}
0 = -\nu\kappa + \Psi_2 + \Phi_{11} - \Lambda
\end{equation}
\begin{equation}
D\lambda = -\nu\kappa^* +\Phi_{20}
\label{D1a}
\end{equation}
\begin{equation}
0 = \sigma\lambda -\nu\kappa + \Psi_2 + 2\Lambda
\end{equation}
\begin{equation}
D\nu = \Psi_3 + \Phi_{21}
\label{D3a}
\end{equation}
\begin{equation}
\Delta\lambda - \delta^*\nu = -\Psi_4
\end{equation}
\begin{equation}
-\delta^*\sigma = - \Psi_1 + \Phi_{01}
\label{D1b}
\end{equation}
\begin{equation}
0 = -\lambda\sigma - \Psi_2 + \Phi_{11}  + \Lambda
\end{equation}
\begin{equation}
\delta\lambda = -\Psi_3 + \Phi_{21}
\label{D2a}
\end{equation}
\begin{equation}
\delta\nu = \lambda\lambda^* + \Phi_{22}
\label{D4a}
\end{equation}
\begin{equation}
0 = -\sigma\nu + \Phi_{12}
\end{equation}
\begin{equation}
-\Delta\sigma = -\kappa\nu^* + \Phi_{02}
\label{D2b}
\end{equation}
\begin{equation}
0 = -\sigma\lambda +\nu\kappa - \Psi_2 - 2\Lambda
\end{equation}
\begin{equation}
0 = - \Psi_3
\end{equation}
\label{NT}
\end{subequations}
These equations give the following solutions for the Ricci and Weyl tensors in terms of the spin coefficients 
\begin{subequations}
\begin{equation}
\Psi_0 = D\sigma - \delta\kappa
\end{equation}
\begin{equation}
\Psi_1 = 0
\end{equation}
\begin{equation}
\Psi_2 = \frac{2}{3}(\nu\kappa - \sigma\lambda)
\end{equation}
\begin{equation}
\Psi_3 = 0
\end{equation}
\begin{equation}
\Psi_4 = \delta^*\nu - \Delta\lambda
\end{equation}
\begin{equation}
\Phi_{00} = -\delta^*\kappa - \sigma\sigma^*
\end{equation}
\begin{equation}
\Phi_{01} = \kappa^*\lambda^*
\end{equation}
\begin{equation}
\Phi_{02} = D\lambda^* + \kappa\nu^*
\end{equation}
\begin{equation}
\Phi_{11} = \frac{\nu\kappa + \sigma\lambda}{2}
\end{equation}
\begin{equation}
\Phi_{12} = \sigma\nu
\end{equation}
\begin{equation}
\Phi_{22} = \delta\nu - \lambda\lambda^*
\end{equation}
\begin{equation}
\Lambda = \frac{\nu\kappa - \sigma\lambda}{6} 
\end{equation}
\label{solutions}
\end{subequations}
\\
Now we try to obtain the four Dirac equations in terms of these eighteen equations (\ref{NT}).
For the first Dirac equation, we add (\ref{D1a}) and (\ref{D1b}), to get
\begin{equation}
D\lambda - \delta^*\sigma = -\nu\kappa^* + \Phi_{20}  + \Phi_{01}
\label{EDNT1}
\end{equation}
Let us compare this with the first of the Dirac equations (\ref{D1}) which can be written as
\begin{equation}
DF_1 +  \delta^{*}F_2 = i\mu_{*} G_1
\end{equation}
Using the correspondence conditions (\ref{match}) and (\ref{vanishspin}) this equation can be written as 
\begin{equation}
D\lambda - \delta^*\sigma = i\mu_{*}\kappa^*
\label{DNT1}
\end{equation}
From (\ref{EDNT1}) and (\ref{DNT1}), we get
\begin{equation}
\Phi_{20}  + \Phi_{01} = (i\mu_* + \nu)\kappa^*
\label{NTC1}
\end{equation}
For the second Dirac equation, we add (\ref{D2a}) and (\ref{D2b}), to get
\begin{equation}
\delta\lambda  -\Delta\sigma=  \Phi_{21} -\kappa\nu^* + \Phi_{02}
\label{EDNT2}
\end{equation}
We compare this equation with the second Dirac equation (\ref{D2}) which can be written as
\begin{equation}
\Delta F_2 + \delta F_1 = i \mu_{*}G_2
\end{equation}
Under (\ref{match}) and (\ref{vanishspin}), this gives
\begin{equation}
\delta \lambda - \Delta\sigma = i\mu_*\nu^*
\label{DNT2}
\end{equation}
From (\ref{EDNT2}) and (\ref{DNT2}), we get
\begin{equation}
\Phi_{21} + \Phi_{02}  = (i\mu_* + \kappa)\nu^* 
\label{NTC2}
\end{equation}
\\
For the third Dirac equation, we add (\ref{D3a}) and (\ref{D3b}), to get
\begin{equation}
D\nu^* - \delta\kappa^* =  \Phi_{12} + \sigma\sigma^* + \Phi_{00}
\label{EDNT3}
\end{equation}
We compare this equation with the third Dirac equation (\ref{D3}) which can be written as
\begin{equation}
DG_2 - \delta G_1 = i\mu_* F_2
\end{equation}
Under (\ref{match}) and (\ref{vanishspin}), this gives
\begin{equation}
D\nu^* - \delta\kappa^* = -i\mu_*\sigma
\label{DNT3}
\end{equation}
From (\ref{EDNT3}) and (\ref{DNT3}), we get
\begin{equation}
\Phi_{12} + \Phi_{00} = -(i\mu_* + \sigma^*)\sigma
\label{NTC3}
\end{equation}
For the fourth Dirac equation we add (\ref{D4a}) and (\ref{D4b}), to get
\begin{equation}
\Delta\kappa^* - \delta^*\nu^* = -(\lambda\lambda^* + \Phi_{22} + \Phi_{10})
\label{EDNT4}
\end{equation}
We compare this equation with the fourth Dirac equation (\ref{D4}) which can be written as
\begin{equation}
\Delta G_1 - \delta^* G_2 = i\mu_{*}F_1
\end{equation}
Under (\ref{match}) and (\ref{vanishspin}), this gives
\begin{equation}
\Delta\kappa^* - \delta^*\nu^* = i\mu_{*}\lambda
\label{DNT4}
\end{equation}
From (\ref{EDNT4}) and (\ref{DNT4}), we get
\begin{equation}
\Phi_{10} + \Phi_{22} = -(i\mu_* + \lambda^*)\lambda
\label{NTC4}
\end{equation}
In summary, the N-P equations (\ref{EDNT1}), (\ref{EDNT2}), (\ref{EDNT3}) and (\ref{EDNT4}) reduce to the four Dirac equations
(\ref{DNT1}), (\ref{DNT2}), (\ref{DNT3}) and (\ref{DNT4}) [which are identical to the Dirac equations 
(\ref{D1}), (\ref{D2}), (\ref{D3}) and (\ref{D4})]
under the correspondence (\ref{match}) and subject to the four conditions (\ref{NTC1}), (\ref{NTC2}), (\ref{NTC3}) and  (\ref{NTC4}).
From the four equations, (\ref{NTC1}), (\ref{NTC2}), (\ref{NTC3}) and  (\ref{NTC4}), we get
\begin{subequations}
\begin{equation}
\Im(\Phi_{01}) = \mu_*\Re(\lambda)
\end{equation}
\begin{equation}
\Im(\Phi_{21}) = \mu_*\Re(\sigma)
\end{equation}
\label{interpret1}
\end{subequations}
\subsection{Constraints on the spin coefficients}
Unfortunately this construction runs into problems. As we know, four of the Ricci components listed in Eqn. (\ref{Ricci}) are real. These reality conditions must be respected by the solutions (\ref{solutions}), and furthermore the eliminant conditions are to be imposed on the Ricci identities since there are only twenty independent Riemann components.  Upon examination we find that this imposes the following undesirable constraints on the four Dirac spinors: 
\begin{subequations}
\begin{equation}
\delta\kappa^* = \delta^*\kappa
\end{equation}
\begin{equation}
\kappa^*\nu^* = \kappa\nu
\end{equation}
\begin{equation}
\lambda^*\sigma^* = \lambda\sigma
\end{equation}
\begin{equation}
\delta^*\nu^* = \delta\nu
\end{equation}
\begin{equation}
\Delta\kappa = \delta^*\sigma = -\kappa^*\lambda^*
\end{equation}
\begin{equation}
D\lambda = -\Delta\sigma^*
\end{equation}
\begin{equation}
D\nu = \delta\lambda = \sigma^*\nu^*
\end{equation}
\end{subequations}
This gives nine constraints on the four non-zero spin coefficients. 
These constraints are unacceptable because the four Dirac spinors are in general completely independent of each other, apart from the normalization constraint, and any other constraint which might arise from 
the intrinsic structure of the Dirac equation.

The origin of these constraints lies in the fact that there are eighteen complex [equivalently thirty-six real] Ricci identities, whereas there are only twenty independent components to the Riemann tensor (See for instance the discussion in \cite{Chandra} Chapter 1.) Now it is known that if one were to include torsion in the theory, the Riemann tensor has thirty-six independent components, and the above mentioned constraints do not arise. This is what we do next: write down the eighteen Ricci identities in the N-P formalism  for a Riemann-Cartan space-time  which includes torsion. We now find that it is possible to establish a correspondence with Dirac equations without having undesirable constraints.

With hindsight, it becomes clear why this correspondence would not work without including some additional structure, such as torsion. The spin coefficients describe the gravitational force, whose role is assumed to diminish in the microscopic limit, in which limit quantum effects are assumed to become more important.  It would be unreasonable to expect these spin coefficients to then take over the role of the Dirac field as the micro-limit is approached. Thus the above costruction should be regarded as a toy model, which does not work, but which makes it easier to understand the following more intricate construction which includes torsion.

\section{Ricci identities for space-time with torsion}
We next consider the Ricci identities in the NP formalism with torsion, since this appears to be a possible way for eliminating the constraints which appear on the spin coefficients and consequently on the Dirac spinors. These identities have been worked out for instance by Jogia and Griffiths \cite{Jogia}. In this case,
the connection is given by
\begin{equation}
\Gamma_{\mu\nu}^{\ \ \lambda} = \left\{ _{\mu\nu}^{\ \ \lambda} \right\} - K_{\mu\nu}^{\ \ \lambda}
\end{equation}
where $\left\{ _{\mu\nu}^{\ \ \lambda} \right\}$ is the Christoffel symbol of the second kind, and  $K_{\mu\nu}^{\ \ \lambda}$ is the contortion tensor which describes the presence of torsion.
In this case, the Ricci tensor is no longer necessarily symmetric, and there are six additional components in the Ricci tensor, described by  the three complex quantities ($\Phi_0, \Phi_1$ and $\Phi_2$).  The Weyl tensor has ten additional components, described by the real quantities ($\Theta_{00}, \Theta_{11}, \Theta_{22}, \chi$) and the complex quantities ($\Theta_{01}, \Theta_{02}, \Theta_{12}$). For definition of these notations please see \cite{Jogia}.

Moreover, the spin coefficients now have an additional term due to torsion, which is denoted by $\gamma_{1}^{nlm}$. Thus, the spin coefficients can now be written as $\gamma_{lmn} = \gamma_{lmn}^{\circ} + K_{nlm} $, where the first term corresponds to the torsion free part and the second term corresponds to the torsion component of the spin coefficients. We further use the following notation to represent the spin-coefficients (see \cite{Jogia} for details):
\begin{equation}
\kappa = \kappa^{\circ} +\kappa_1, \qquad \rho = \rho^{\circ} + \rho_1, \ \ \ \ \ \ \text{etc}.
\end{equation}
The Ricci identities are modified when torsion is included, and are now given by \cite{Jogia}
\begin{subequations}
\begin{align}
D\rho - \delta^{*}\kappa = &\ \rho (\rho + \epsilon + \epsilon^{*}) + \sigma \sigma^{*} - \tau \kappa^{*} - \kappa (3\alpha + \beta^{*} - \pi) + \Phi_{00} \nonumber  \\
&- \rho(\rho_1 - \epsilon_1 + \epsilon_{1}^{*} ) - \sigma \sigma_1^{*} + \tau \kappa_1^{*} + \kappa(\alpha_1 + \beta_{1}^{*} - \pi_1 ) + i\Theta_{00}\\ \nonumber \\
D\sigma - \delta\kappa = &\ \rho\sigma +\sigma(\rho^* + 3\epsilon - \epsilon^*) -\tau\kappa - \kappa(\alpha^* + 3\beta - \pi^*) + \Psi_0 \nonumber \\
& -\rho\sigma_1 -\sigma(\rho_{1}^{*} + \epsilon_1 - \epsilon_{1}^{*}) + \tau\kappa_1 + \kappa(\alpha_{1}^{*} + \beta_1 -\pi_{1}^{*})\\ \nonumber \\
D\tau - \Delta\kappa  = &\ \rho(\tau +\pi^*) + \sigma(\tau^* + \pi) - \tau(\epsilon^* - \epsilon) - \kappa(3\gamma + \gamma^*) + \Psi_1 + \Phi_{01} -\rho(\tau_1 + \pi_{1}^{*}) \nonumber \\
& - \sigma(\tau_{1}^{*} +\pi_1) + \tau(\epsilon_{1}^{*} + \epsilon_1) + \kappa(\gamma_1 + \gamma_{1}^{*}) + i\Theta_{01} + \Phi_0  \\ \nonumber \\
D\alpha - \delta^{*} \epsilon = &\ \alpha (\rho + \epsilon^* - \epsilon) + \beta \sigma^* - \gamma \kappa^* - \epsilon(\alpha + \beta^* - \pi) + \rho \pi - \kappa \lambda + \Phi_{10} \nonumber \\
& -\alpha (\rho_1 + \epsilon_{1}^{*} - \epsilon_1 ) -\beta \sigma_{1}^{*}  + \gamma\kappa_{1}^{*} + \epsilon(\alpha_{1} + \beta_{1}^{*} - \pi_1 ) + i\Theta_{10}  \\  \nonumber \\
D\beta - \delta\epsilon = & \ \alpha \sigma + \beta( \rho^* - \epsilon^*) - \gamma\kappa - \epsilon(\alpha^* - \pi^*) + \sigma\pi - \kappa\mu + \Psi_1 \nonumber \\
& - \alpha\sigma_1 - \beta( \rho_{1}^{*} - \epsilon_{1}^{*} + \epsilon_1 ) + \gamma\kappa_1 + \epsilon( \alpha_{1}^{*} + \beta_1 - \pi_{1}^{*} ) - \Phi_0 \\ \nonumber \\
D\gamma - \Delta\epsilon = &\ \alpha(\tau + \pi^*) + \beta(\tau^* + \pi) - \gamma(\epsilon + \epsilon^*) - \epsilon(\gamma + \gamma^*) + \tau\pi - \kappa\nu + \Psi_2 - \Lambda + \Phi_{11}\nonumber \\
& - \alpha (\tau_1 + \pi_{1}^{*}) - \beta( \tau_{1}^{*} + \pi_1) + \gamma(\epsilon_1 + \epsilon_{1}^{*}) + \epsilon(\gamma_1 + \gamma_{1}^{*}) + i\Theta_{11} -i\chi \\  \nonumber \\
D\lambda - \delta^* \pi = & \ \mu\sigma^* + \lambda(\rho - 3\epsilon + \epsilon^*) - \pi(-\alpha + \beta^* - \pi) - \nu\kappa^* + \Phi_{20} \nonumber\\
& - \mu\sigma_{1}^{*} - \lambda(\rho_1 - \epsilon_1 + \epsilon_{1}^{*}) + \pi(\alpha_1 + \beta_{1}^{*} - \pi_1 ) + \nu\kappa_{1}^{*} + i\Theta_{20} \\  \nonumber \\
D\mu - \delta\pi = & \ \mu(\rho^* - \epsilon - \epsilon^* ) + \lambda\sigma - \pi (\alpha^* - \beta - \pi^*) - \nu\kappa + \Psi_2 + 2\Lambda  \nonumber \\
& - \mu(\rho_{1}^{*}  + \epsilon_1 - \epsilon_{1}^{*} ) - \lambda\sigma_1 + \pi ( \alpha_{1}^{*} + \beta_1 - \pi_{1}^{*}) + \nu\kappa_1 - \Phi_1  \\ \nonumber  \\
D\nu - \Delta\pi = & \ \mu(\pi + \tau^*) + \lambda ( \pi^* + \tau ) - \pi(\gamma^* - \gamma) - \nu(3\epsilon + \epsilon^*) + \Psi_3 + \Phi_{21} \nonumber \\
& - \mu(\pi_1 + \tau_{1}^{*} ) - \lambda(\pi_{1}^{*} + \tau_1 ) + \pi(\gamma_{1}^{*} + \gamma_1) + \nu(\epsilon_1 + \epsilon_{1}^{*} ) + i\Theta_{21} - \Phi_2 \\ \nonumber \\
\Delta\lambda - \delta^* \nu =&\ - \mu\lambda  - \lambda (\mu^* + 3\gamma -\gamma^*) + \pi\nu + \nu (3\alpha + \beta^* - \tau^*) - \Psi_4 \nonumber\\
& + \mu\lambda_1 + \lambda(\mu_{1}^{*} + \gamma_1 - \gamma_{1}^{*} ) - \pi\nu_1 - \nu(\alpha_1 +\beta_{1}^{*}  - \tau_{1}^{*})  \\ \nonumber\\
\delta\rho - \delta^* \sigma = & \rho( \alpha^* + \beta) + \sigma(\beta^* - 3\alpha ) + \tau(\rho - \rho^*) + \kappa(\mu - \mu^*) - \Psi_1 + \Phi_{01} \nonumber \\ 
& -\rho(\alpha_{1}^{*} - \beta_1 ) - \sigma(\beta_{1}^{*} - \alpha_1) - \tau(\rho_1 - \rho_{1}^{*} ) - \kappa(\mu_1 - \mu_{1}^{*} ) + i\Theta_{01} - \Phi_0  \\ \nonumber \\
\delta\alpha - \delta^* \beta = & \alpha (\alpha^* - \beta) + \beta (\beta^* - \alpha) + \gamma( \rho - \rho^*) + \epsilon ( \mu - \mu^*) + \rho\mu - \sigma\lambda- \Psi_2 + \Lambda + \Phi_{11} \nonumber \\
& - \alpha (\alpha_{1}^{*} - \beta_1)  - \beta(\beta_{1}^{*} - \alpha_1) - \gamma(\rho_1 - \rho_{1}^{*} ) - \epsilon(\mu_1 - \mu_{1}^{*})+ i\Theta_{11} + i\chi \\ \nonumber \\
\delta\lambda - \delta^* \mu = & \mu (\beta^* + \alpha) + \lambda(\alpha^* - 3\beta) + \pi(\mu - \mu^*) + \nu(\rho - \rho^*) - \Psi_3 + \Phi_{21} \nonumber \\
& - \mu(\beta_{1}^{*} - \alpha_1) - \lambda(\alpha_{1}^{*} - \beta_1) - \pi(\mu_1 - \mu_{1}^{*}) - \nu(\rho_1 - \rho_{1}^{*}) + i\Theta_{21} + \Phi_2 \\ \nonumber \\
\delta\nu - \Delta\mu =& \  \mu(\mu +\gamma+\gamma^*) + \lambda\lambda^* - \pi\nu^* - \nu(3\beta + \alpha^*  - \tau)  + \Phi_{22} \nonumber \\
& - \mu(\mu_{1} - \gamma_1 + \gamma_{1}^{*} ) - \lambda\lambda_{1}^{*} + \pi \nu_{1}^{*}  + \nu(\beta_1 + \alpha_{1}^{*} - \tau_1) + i\Theta_{22} \\ \nonumber \\
\delta\gamma - \Delta\beta = & \alpha \lambda^* + \beta(\mu - \gamma + \gamma^*)  - \gamma(\beta+ \alpha^* - \tau) - \epsilon\nu^* + \mu\tau - \nu\sigma + \Phi_{12} \nonumber \\
& - \alpha\lambda_{1}^* - \beta(\mu_1 - \gamma_1 + \gamma_{1}^*) + \gamma ( \beta_1 + \alpha_{1}^* - \tau_1) + \epsilon\nu_{1}^* + i\Theta_{12}  \\ \nonumber \\
\delta \tau - \Delta\sigma = & \rho \lambda^* + \sigma(\mu - 3\gamma +\gamma^*) - \tau (\alpha^* - \beta - \tau ) - \kappa\nu^* + + \Phi_{02} \nonumber \\
& - \rho \lambda_{1}^*  - \sigma (\mu_1 - \gamma_1 + \gamma_{1}^* ) + \tau (\alpha_{1}^* + \beta_1 -\tau_ 1) + \kappa\nu_{1}^* + i\Theta_{02} \\ \nonumber \\
\Delta\rho - \delta^* \tau = & \ -\rho(\mu^* - \gamma - \gamma^* ) - \sigma \lambda + \tau (\beta^* - \alpha - \tau^* ) + \kappa\nu - \Psi_2 - 2\Lambda \nonumber \\
& + \rho(\mu_{1}^* + \gamma_1 - \gamma_{1}^*) + \sigma\lambda_1 - \tau(\beta_{1}^* + \alpha_1 - \tau_{1}^*) - \kappa\nu_1 - \Phi_1 \\ \nonumber \\
\Delta\alpha - \delta^* \gamma = & \ - \alpha(\mu^* - \gamma^*) - \beta\lambda + \gamma(\beta^* - \tau^*) + \epsilon\nu + \rho\nu - \tau\lambda - \Psi_3 \nonumber \\
& + \alpha (\mu_{1}^* + \gamma_1 - \gamma_{1}^*) + \beta\lambda_1 - \gamma(\beta_{1}^* + \alpha_1 - \tau_{1}^*) - \epsilon\nu_1 - \Phi_2
\end{align}
\label{torsionnp}
\end{subequations}
These generalize the Ricci identities given previously in the torsion-free case, and as may be verified, reduce to those equations when torsion is set to zero. 

Once again, we use (\ref{vanishspin}) to retain only four non-zero spin-coefficients. Thus, equation (\ref{torsionnp}) gives    
\begin{subequations}
\begin{equation}
-\delta^* \kappa = \sigma \sigma^{*} + \Phi_{00} - \sigma\sigma_{1}^{*} + i\Theta_{00}
\end{equation}
\begin{equation}
D\sigma - \delta\kappa = \Psi_0 
\end{equation}
\begin{equation}
-\Delta\kappa = \Psi_1 + \Phi_{01} + i\Theta_{01} + \Phi_0  
\end{equation}
\begin{equation}
0 = -\kappa\lambda + \Phi_{10} + i\Theta_{10}
\end{equation}
\begin{equation}
0 = \Psi_1 - \Phi_0
\end{equation}
\begin{equation}
0 = -\kappa\nu + \Psi_2 - \Lambda + \Phi_{11} + i\Theta_{11} - i\chi
\end{equation}
\begin{equation}
D\lambda = -\nu\kappa^* + \Phi_{20} + \nu\kappa_{1}^* + i\Theta_{20}
\end{equation}
\begin{equation}
0 = \lambda\sigma -\nu\kappa + \Psi_2 + 2\Lambda - \lambda\sigma_1 + \nu\kappa_1 - \Phi_1
\end{equation}
\begin{equation}
D\nu = \Psi_3 + \Phi_{21}  + i\Theta_{21} - \Phi_2
\end{equation}
\begin{equation}
\Delta \lambda - \delta^* \nu = -\Psi_4 
\end{equation}
\begin{equation}
-\delta^* \sigma = -\Psi_1 + \Phi_{01} + i\Theta_{01} - \Phi_0
\end{equation}
\begin{equation}
0 = -\sigma\lambda -\Psi_2 + \Lambda + \Phi_{11} +i\Theta_{11} +i\chi
\end{equation}
\begin{equation}
\delta\lambda =  -\Psi_3 + \Phi_{21} + i\Theta_{21} + \Phi_2
\end{equation}
\begin{equation}
\delta\nu = \lambda\lambda^* + \Phi_{22} - \lambda\lambda_{1}^* + i\Theta_{22} 
\end{equation}
\begin{equation}
0 = -\nu\sigma + \Phi_{12} + i\Theta_{12}
\end{equation}
\begin{equation}
-\Delta\sigma = -\kappa\nu^* + \Phi_{02} + \kappa\nu_{1}^* + i\Theta_{02}
\end{equation}
\begin{equation}
0 = -\sigma\lambda + \kappa\nu - \Psi_2 - 2\Lambda + \sigma\lambda_1 - \kappa\nu_1 - \Phi_1
\end{equation}
\begin{equation}
\Psi_3 = - \Phi_2
\end{equation}
\end{subequations}
Solving these equations for the Ricci and the Weyl tensors in terms of the spin coefficients gives the following results
\begin{subequations}
\begin{equation}
\Psi_0 = D\sigma - \delta\kappa
\end{equation}
\begin{equation}
\Psi_1 = \frac{\delta^* \sigma - \Delta\kappa}{4}
\end{equation}
\begin{equation}
\Psi_2 = \frac{5\nu\kappa - \nu^*\kappa^* - 5\lambda\sigma + \lambda^*\sigma^*}{6} + \frac{2\lambda\sigma_1 - \lambda^*\sigma_{1}^* + 2\lambda_{1}\sigma - \lambda_{1}^*\sigma^*}{6}
- \frac{2\nu\kappa_1 - \nu^*\kappa_{1}^* + 2\nu_1\kappa - \nu_{1}^*\kappa^*}{6}
\end{equation}
\begin{equation}
\Psi_3 = \frac{D\nu -\delta\lambda}{4}
\end{equation}
\begin{equation}
\Psi_4 = \delta^* \nu - \Delta \lambda
\end{equation}
\begin{equation}
\Phi_{00} = -\sigma\sigma^* - \frac{\delta\kappa^* + \delta^*\kappa}{2} + \frac{\sigma\sigma_{1}^* + \sigma^*\sigma_1}{2} = -\sigma\sigma^*  -\Re(\delta^*\kappa) + \Re(\sigma\sigma_{1}^*)
\end{equation}
\begin{equation}
\Phi_{01} = \frac{\kappa^*\lambda^*}{2} - \frac{\delta^*\sigma + \Delta\kappa}{4}
\end{equation}
\begin{equation}
\Phi_{02} = \frac{D\lambda^* - \Delta\sigma}{2} + \kappa\nu^*   - \frac{\kappa\nu_{1}^* + \kappa_1\nu^*}{2}
\end{equation}
\begin{equation}
\Phi_{11} = \frac{\kappa\nu + \kappa^*\nu^* + \sigma\lambda + \sigma^*\lambda^*}{4} = \frac{\Re(\kappa\nu) + \Re(\sigma\lambda)}{2}
\end{equation}
\begin{equation}
\Phi_{12} = \frac{D\nu^* + \delta^*\lambda^*}{4} + \frac{\sigma\nu}{2}
\end{equation}
\begin{equation}
\Phi_{22} = - \lambda\lambda^*+ \frac{\lambda\lambda_{1}^* + \lambda^*\lambda_1}{2} +\frac{\delta\nu + \delta^*\nu^*}{2} = -\lambda\lambda^* + \Re(\lambda\lambda_{1}^*) + \Re(\delta\nu)
\end{equation}
\begin{equation}
\begin{split}
\Lambda &= \frac{\kappa\nu + \kappa^*\nu^*- \lambda\sigma - \lambda^*\sigma^*}{12}  + \frac{\lambda\sigma_1 + \lambda^*\sigma_{1}^* + \lambda_1\sigma + \lambda_{1}^*\sigma^*}{12} - \frac{\nu\kappa_1 + \nu^*\kappa_{1}^* + \nu_1\kappa + \nu_{1}^*\kappa^*}{12} \\
&= \frac{\Re(\kappa\nu)  - \Re(\nu\kappa_1) - \Re(\nu_1\kappa) - \Re(\sigma\lambda) + \Re(\lambda\sigma_1) + \Re(\lambda_1\sigma)}{6}
\end{split}
\end{equation}
\begin{equation}
\Theta_{00} = \frac{i}{2}(\sigma^*\sigma_1 - \sigma\sigma_{1}^*) - \frac{i}{2}(\delta\kappa^* - \delta^*\kappa) = \Im(\sigma\sigma_{1}^*)  - \Im(\delta^*\kappa)
\end{equation}
\begin{equation}
\Theta_{01} = \frac{i}{2} \kappa^*\lambda^* + \frac{i}{4}( \delta^*\sigma + \Delta\kappa)
\end{equation}
\begin{equation}
\Theta_{02} = \frac{i}{2}(D\lambda^* +\Delta\sigma) - \frac{i}{2} ( \kappa_1\nu^* - \kappa\nu_{1}^*)
\end{equation}
\begin{equation}
\Theta_{11} = -\frac{i}{4} (\kappa\nu - \kappa^*\nu^*) - \frac{i}{4}(\sigma\lambda - \sigma^*\lambda^*) = \frac{\Im(\kappa\nu) + \Im(\sigma\lambda)}{2}
\end{equation}
\begin{equation}
\Theta_{12} = \frac{i}{4} (D\nu^* + \delta^*\lambda^*) - \frac{i}{2}\sigma\nu
\end{equation}
\begin{equation}
\Theta_{22} = -\frac{i}{2} (\delta\nu -\delta^*\nu^*) - \frac{i}{2}(\lambda\lambda_{1}^* - \lambda^*\lambda_1) = \Im(\delta\nu) + \Im(\lambda\lambda_{1}^*)
\end{equation}
\begin{equation}
\chi = -\frac{i}{4}(\nu\kappa - \nu^*\kappa^* - \sigma\lambda + \sigma^*\lambda^*)\ - \ \frac{i}{4}(\lambda\sigma_1 - \lambda^*\sigma_{1}^* + \lambda_1\sigma - \lambda_{1}^*\sigma^*)\ + \  \frac{i}{4}(\nu\kappa_1 - \nu^*\kappa_{1}^* + \kappa\nu_1 - \kappa^*\nu_{1}^*) 
\end{equation}
\begin{equation}
\Phi_0 = \frac{\delta^*\sigma - \Delta\kappa}{4}
\end{equation}
\begin{equation}
\Phi_1 = \frac{\sigma\lambda_1 - \sigma_1\lambda}{2} + \frac{ \kappa_1\nu - \kappa\nu_1}{2}
\end{equation}
\begin{equation}
\Phi_2 = \frac{\delta\lambda - D\nu}{4}
\end{equation}
\label{torsion}
\end{subequations}
It can be verified that now there are no constraints imposed on the four spin-coefficients because of the reality conditions of the Ricci and Weyl tensor components. The introduction of torsion has made a significant difference.

As before, using (\ref{match}) for making correspondence with the spinor components of  the Dirac equations and using the same process as used from equation (\ref{EDNT1}) to (\ref{NTC4}) we get the following four conditions under which Dirac equations are contained within the Ricci identities
\begin{equation}
\Phi_{20} + i\Theta_{20} + \Phi_{01} + i\Theta_{01} - \Psi_1 - \Phi_0  = (i\mu_* +\nu)\kappa^* - \nu\kappa_{1}^*
\end{equation}
\begin{equation}
\Phi_{21} + i\Theta_{21} + \Phi_2 - \Psi_3 + \Phi_{02} + i\Theta_{02}  = (i\mu_* + \kappa)\nu^* - \kappa\nu_{1}^*
\end{equation}
\begin{equation}
i\Theta_{12} - \Phi_{12} + i\Theta_{00} - \Phi_{00} + \Phi_{2}^* -\Psi_{3}^* = (i\mu_* + \sigma^*)\sigma - \sigma^*\sigma_1
\end{equation}
\begin{equation}
i\Theta_{10} - \Phi_{10} - \Phi_{0}^* - \Psi_{1}^* + i\Theta_{22} - \Phi_{22} = (i\mu_* + \lambda^*)\lambda - \lambda^*\lambda_1
\end{equation}
%However, in this case, $\Lambda$ is not real and thus that would lead to a constraint on the spin coefficients and since we are co-relating the spin coefficients with the Dirac spinors, this would lead to a constraint on the Dirac spinors, i.e., $\Lambda = \Lambda^*$.

Here, an important remark is in order. We have made correspondence with the torsion-free Dirac equations written in the N-P formalism: these are the same as the four equations (\ref{D1}) to (\ref{D4}). It is known (see for instance \cite{Zecca}), \cite{Hehl}) that in the presence of torsion the Dirac equation acquires 
quadratic non-linear terms dependent on the wave-function. At present we make the assumption that these quadratic terms maybe neglected, so that we recover standard flat space-time Dirac equations. In a future work we will address the possible significance of these non-linear terms.

If we take the torsion free part of the four spin-coefficients ($\kappa, \lambda, \sigma$ and $\nu$) to be zero, i.e., $ \kappa = \kappa_1, \sigma = \sigma_1, \nu = \nu_1$ and $\lambda = \lambda_1$ we still do not get any constraint on the spin-coefficients. In that case, $\Psi_0,\ \Psi_1,\ \Psi_3,\ \Psi_4,\ \Phi_0,\ \Phi_2, \ \Phi_{01},\ \Phi_{11},\ \Phi_{12},\ \Theta_{01},\ \Theta_{11},\ \Theta_{12}$  are unchanged, i.e., they are given by (\ref{torsion}). The other Riemann components change, and are given by
\begin{subequations}
\begin{equation}
\Psi_2 = \frac{\kappa\nu + \kappa^*\nu^* - \sigma\lambda - \sigma^*\lambda^*}{6} = \frac{\Re(\kappa\nu) - \Re(\sigma\lambda)}{3}
\end{equation}
\begin{equation}
\Phi_{00} = - \frac{\delta\kappa^* + \delta^*\kappa}{2} = -\Re(\delta\kappa^*)
\end{equation}
\begin{equation}
\Phi_{02} = \frac{D\lambda^* - \Delta\sigma}{2}
\end{equation}
\begin{equation}
\Phi_{22} = \frac{\delta\nu + \delta^*\nu^*}{2} = \Re(\delta\nu)
\end{equation}
\begin{equation}
\Lambda = \frac{\sigma\lambda + \sigma^*\lambda^* - \kappa\nu - \kappa^*\nu^*}{12} = \frac{\Re(\sigma\lambda) - \Re(\kappa\nu)}{6}
\end{equation}
\begin{equation}
\Theta_{00} = \frac{i}{2}(\delta^*\kappa - \delta\kappa^*) = \Im(\delta\kappa^*)
\end{equation}
\begin{equation}
\Theta_{02} = \frac{i}{2}(D\lambda^* + \Delta\sigma)
\end{equation}
\begin{equation}
\Theta_{22} = \frac{i}{2} (\delta^*\nu^* - \delta\nu) = \Im(\delta\nu)
\end{equation}
\begin{equation}
\Phi_1 = 0
\end{equation}
\begin{equation}
\chi = \frac{i}{4} (\kappa\nu - \kappa^*\nu^*) - \frac{i}{4}(\sigma\lambda - \sigma^*\lambda^*) = \frac{\Im(\sigma\lambda) -\Im(\kappa\nu)}{2}
\end{equation}
\label{notorsion}
\end{subequations}
In this case, the four conditions under which the Ricci identities reduce to Dirac equations are
\begin{equation}
\Phi_{20} + i\Theta_{20} + \Phi_{01} + i\Theta_{01} - \Psi_1 - \Phi_0  = i\mu_*\kappa^* 
\label{dir1}
\end{equation}
\begin{equation}
\Phi_{21} + i\Theta_{21} + \Phi_2 - \Psi_3 + \Phi_{02} + i\Theta_{02}  = i\mu_* \nu^* 
\label{dir2}
\end{equation}
\begin{equation}
i\Theta_{12} - \Phi_{12} + i\Theta_{00} - \Phi_{00} + \Phi_{2}^* -\Psi_{3}^* = i\mu_* \sigma 
\label{dir3}
\end{equation}
\begin{equation}
i\Theta_{10} - \Phi_{10} - \Phi_{0}^* - \Psi_{1}^* + i\Theta_{22} - \Phi_{22} = i\mu_* \lambda 
\label{dir4}
\end{equation}

It is this case, which we call the torsion dominated limit, which is of central interest to us. We are proposing that the Dirac quantum state be taken as being proportional to  the torsion, with the latter being expressed in terms of the complex spin-coefficient (the antisymmetric part of the connection). In this limit, the symmetric part of the connection is zero - this may be interpreted as the zero gravity limit; a flat spacetime Minkowski metric for which the Christoffel symbols vanish. We do not at this stage know how to disentangle the field equations for the explicit Ricci and Weyl components from these equations. [The opposite extreme would supposedly be the gravity dominated limit (Einstein gravity), in which the torsion is not zero, but is spatially localized, corresponding to gravity induced localization of the wave-function. However, this remains to be proved.] In terms of the tetrad components of the Riemann tensor, the above four equations can be written as (where the subscripts 1, 2, 3 and 4 stand for the tetrads $\bf l, n, m, \overline{m}$ respectively)
\begin{equation}
R_{1213} +R_{1312} +R_{4142} - R_{3134} = i\mu_* K_{141}
\end{equation}
\begin{equation}
R_{2123} +R_{2321} +R_{4241} - R_{3234} = i\mu_* K_{242}
\end{equation}
\begin{equation}
R_{4342} +R_{4243} +R_{1413} - R_{2421} = i\mu_* K_{414}
\end{equation}
\begin{equation}
R_{4341} +R_{4143} +R_{2423} - R_{1412} = i\mu_* K_{424}
\end{equation}
where $K_{\mu\nu\lambda} $is the contortion tensor. The explicit relation between the Riemann tensor and the N-P scalars can be found in Eqns. (3.10) and (3.11) of \cite{Jogia}. In order that these equations be written in covariant tensor notation, it should turn out that the tetrad components drop out upon transformation to the tensor form. That does not seem to happen: one possible explanation is that the introduction of a complex torsion makes the N-P description more fundamental, and inequivalent to the tensor description, with equivalence being restored only for real torsion.

The above four equations are equivalent to the four Dirac equations. The solutions of the Dirac equations can be thought of as providing solutions for the four non-zero torsion components, which in turn determine the solution for the Riemann tensor. A significant departure from standard theories with torsion is that here the torsion is not real, but complex! We will comment below on how one might possibly interpret this.

A remark on the Bianchi identities for torsion and curvature, as listed in detail for instance in \cite{Jogia}. 
We have explicitly verified that when the torsion free part is zero, the Bianchi identities continue to hold.
This is to be expected, because the solutions for the Riemann tensor that we have constructed follow from the Bianchi identities. Given the peculiar structure of the above four equations (\ref{dir1})-(\ref{dir4}), the Bianchi identities do not yield constraints on the Dirac equations. For similar reasons, we expect the Bianchi identities to continue to hold when the torsion-free part is non-zero.

\section{Critical discussion and possible interpretation}
We started by suggesting that there should be a common description for the gravitational field and the Dirac field of a relativistic particle, for both could be thought of as being sourced by the particle'€™s mass $m$. One limit is expected to arise when $m\gg m_{pl}$ and $R_S \gg L$ and the other limit is supposed to arise when $m\ll m_{pl}$ and $R_S \ll L$. We first tried to arrive at this common description by using the Newman-Penrose formalism and the Ricci identities for a Riemmanian spacetime, but we found that the idea does not work. There are too few independent components of the Riemann tensor, and the resulting constraints on the Ricci identities translate into undesired constraints on the four non-zero spin coefficients (equivalently the four Dirac spinors). In so far as the unusual assumed  correspondence between the four non-vanishing spin-coefficients and the four Dirac spinors is concerned, we cannot think of any mathematical / physical inconsistency in doing so.  Nor, it should be said, have we provided an explicit convincing argument for doing so, except that it seems to help achieve the purpose set out in the introduction.

We next considered the case of a Riemann-Cartan spacetime, possessed with a non-zero torsion, for it allows thirtysix independent components for the Riemann tensor, and hence no constraints coming from the Ricci identities. Motivation for including torsion comes also from the fact that it makes Einstein gravity more in line with the Dirac theory: in quantum theory [Dirac equation] elementary particles are represented by irreducible representations of the Poincare group (not the Lorentz group) labeled by mass and spin. In Einstein gravity, the structure group is the Lorentz group, not the Poincare group: there is no room for translations. The inclusion of translations gives rise to torsion, and to a Riemann-Cartan spacetime.

It is to be noted that in the four Ricci identities (\ref{dir1}) to (\ref{dir4}) given above, the source term is complex. In general, this will amount to a complex contribution to the Riemann tensor [Minkowski flat spacetime, with contributions to a complex Riemann coming from torsion]. This possibly suggests that a description in terms of the complex spin coefficients is more fundamental, compared to a description in terms of the connection, until the macroscopic classical limit is approached, when torsion would be localized, and the matter free region would again have a real Riemann tensor.

There next comes the important question of what form the field equations will take, when torsion and gravity (the symmetric part of the connection) are both present. We do not have the final answer to this, but taking cue from the schematic form of the action (\ref{schematicaction}) we symbolically suggest
\begin{equation}
\frac{c^3}{G} G_{\mu\nu} - \frac{i\hbar}{l^2} [Riem]^{torsion}_{\mu\nu} = \frac{mc}{l^2} f_{\mu\nu}
\label{genee}
\end{equation}
Here, $l$ is a length scale to be determined by the solution of the field equations.
In the first term on the left, $G_{\mu\nu}$ is the standard Einstein tensor, and if this term dominates over the second term, the right hand side should reduce to the symmetric stress-energy tensor: the quantities $f_{\mu\nu}$ should define an inverse length scale $1/l$, so that the right hand side
reduces to a mass density times speed of light. It is seen by inspection that when the second term on the left is negligible, the length scale $l$ is inevitably the Schwarzschild radius  $R_S$.

The second term on the left is symbolically the new idea we have tried to introduce in this paper, namely that the Dirac field might be identified with a complex torsion in a Riemann-Cartan spacetime. If this term dominates over the first term on the left hand side, then by the quantities $[Riem]^{torsion}_{\mu\nu}$ we mean the information contained on the left hand side of the four equations (\ref{dir1}) to (\ref{dir4}) above. In this case, the quantities $f_{\mu\nu}$ on the right hand should be linearly proportional to the appropriate spin coefficients, so that the Dirac equations are recovered. If the spin coefficient is a quantity of the order $1/l$, then on inspection the scale $l$ is found to be of the order of Compton wavelength $L$,  as expected. Again one can argue that if the first term dominates we have $R_S\gg L$ and $m\gg m_{pl}$ and if the second term dominates the inequalities are reversed; these arguments are of the same nature as those below Eqn. (\ref{schematicaction}). 

When $m\sim m_{pl}$ we expect both terms on the left to be present, and this is the essence of our argument: to think of the Dirac equations as resulting from a modified Einstein gravity, with the modification being brought about by inclusion of a complex torsion on a Riemann-Cartan spacetime, via the Newman-Penrose equations. Here we could think of the first term, which contains gravity, as introducing self-gravity into the formalism, which is the sought for modification we are after, in principle. We know that the non-relativistic limit of the Dirac equation is the Schr\"odinger equation. If we could construct the non-relativistic limit of the gravity-Dirac equation (\ref{genee}) we expect to arrive at a non-linear Schr\"odinger equation which incorporates the effect of self-gravity, and which could be of assistance in understanding gravity induced collapse of the wave-function. We hope to pursue this analysis in the near future.

How do we interpret the dynamics of the theory symbolically described by Eqn. (\ref{genee})? First, we must say how the right hand side - the matter source - is constructed from the Dirac field (now interpreted as a complex torsion) and other matter fields.  In analogy with torsion theories of gravity, one constructs a matter Lagrangian density which depends on the Dirac field (now identified with torsion), on any additional matter fields to be included, on  the metric (or more fundamentally, the tetrad) and on the torsion. [See for instance Section II.E of \cite{Hehl} for a motivating discussion]. Variation w.r.t. the metric / tetrad yields, instead of the conventional Einstein equations relating the Einstein tensor to the energy-momentum tensor, a field eqn. such as Eqn. (\ref{genee}), because the geometric part of the action also depends on the torsion [the geometric part being the analog  of the first two terms in the schematic action (\ref{schematicaction})]. The dynamics is understood as a solution of these field equations for the spin connection, given the matter distribution. The symmetric part of the connection can then be inverted to obtain the tetrads, from which the metric may be constructed. The torsion part of the connection then defines the Dirac fields. This seems to provide a useful way to think of the gravitational field and the Dirac fields as the symmetric and antisymmetric parts of the connection (the latter being complex) respectively.

Mention must be made of the very elegant but much neglected Einstein-Cartan-Sciama-Kibble theory which generalizes Einstein gravity to include torsion (for a review see \cite{Hehl}). The basic principles on which the ECSK theory is built are very sound. The symmetry group of the Minkowski spacetime in special relativity is the Poincar\'e group, which includes both Lorentz transformations and translations. This gives rise to conservation laws of angular momentum and energy momentum. In general relativity, which describes a curved spacetime, the structure group which acts on tangent spaces is the Lorentz group, {\it not} the Poincare group. In GR there is no room for translations. By introducing torsion and relating it to the intrinsic (spin) angular momentum, Cartan showed that translations are brought in, and the structure group of the spacetime is now the Poincare group, not the Lorentz group. Curvature is related to Lorentz transformations in the same way that torsion is related to translations. \cite{Trautman}

On the other hand, in quantum theory, elementary particles correspond to irreducible, unitary representations of the Poincare group (not the Lorentz group), and these are labeled by mass as well as spin. It then seems likely that if a unifying description between classical gravity and quantum theory is to be looked for, the ECSK theory is a more plausible classical candidate, as compared to Einstein's general relativity. Why then is it still the case that not much attention has been paid to the ECSK theory, when it comes to exploring quantum gravity? One possible reason of course is that there is little evidence for torsion in the observed astrophysical world, so it is hard to take it seriously. Also, outside of matter, torsion vanishes, and in empty space the ECSK theory coincides with Einstein gravity.

It should be noted that strictly speaking, the ECSK theory is a special case of a more general gauge theory, the Poincar\'e gauge theory of gravity [PG], which is obtained by gauging the Poincar\'e group \cite{Blagojevic, Hehl3, Hehl1, Hehl2}. The ECSK theory is a special case of PG, obtained when the gravitational Lagrangian is chosen to be the Einstein-Hilbert Lagrangian. It is interesting that the special case of PG, obtained when the Ricci tensor is set to zero, has been well-studied in the literature, and is known as teleparallel gravity [TG] (torsion without gravity) [see for instance Fig. 2  of \cite{Hehl1}]. TG comes close to our representation of Dirac equations in terms of a complex torsion, and it will be significant to explore further the relation of our work to TG. Moreover, working backwards, we could ask what impact the complex nature of the torsion has on the choice of the symmetry group one started from, i.e. the Poincar\'e group P(3,1).  

 The points made  above seem to provide us encouragement for the idea we have suggested in this article. If torsion is theoretically important (restores Poincare group in a curved spacetime, bringing parity with quantum theory) but not important observationally, why not make torsion complex, and identify torsion with the Dirac quantum state?  If that were to be done, the ECSK theory could possibly become a bridge between Einstein gravity on the one hand, and the Dirac equation on the other: the former being the gravity dominated limit, and the latter being the torsion dominated limit. It is then also understandable that torsion vanishes outside the matter source: matter is only where the quantum state (torsion) is. It seems worthwhile to explore the ECSK theory from this point of view, and to investigate if this theory could assist in understanding gravity induced collapse of the wave-function. 
 
We also note that the form of equations (\ref{torsion}) and (\ref{notorsion}) is very similar to the extra terms that enter the energy-momentum tensor due to the presence of torsion \cite{Trautman}. The extra terms in the energy-momentum tensor due to torsion are either the products of two spin coefficients or the derivatives of spin coefficients which is the same structure as the terms in equations (\ref{torsion}) and (\ref{notorsion}). This suggests some similarity between the spin coefficients and spin, since in the first case the extra terms are the spin tensor components and in the second case they are the spin coefficients. Moreover, the spin coefficients are anti-symmetric in the first two indices just like the spin-tensor components.
 
 Another motivation for considering gravity as a mediating process in wave-function collapse is the following premise. The gravitational field (i.e. the Coulomb part, not the gravitational wave) is inseparable from its matter source. It is hence reasonable to require that any physical process which explains dynamical collapse and localization of the material particle should also simultaneously account for its accompanying classical gravity field. Thus the involvement of the `quantum gravitational field` of the quantum object, as it approaches the macroscopic regime, is strongly indicated in the localization process. This also suggests that quantum gravity and the quantum measurement problem have a lot to do with each other. One could even push the argument further and note that since wave-function collapse is a non-linear process, some sort of non-linearity should be inherent in a quantum theory of gravity as well.

\bigskip

\noindent{\bf Acknowledgement:} This work is supported by a grant from the John Templeton Foundation (\# 39530).

\bigskip

\centerline{\bf REFERENCES}

\bibliography{biblioqmts3}

\end{document}